%% file: conference_IEEE.tex
\documentclass[conference]{IEEEtran}
\IEEEoverridecommandlockouts
\usepackage{cite}
\usepackage{amsmath,amssymb,amsfonts}
\usepackage{algorithmic}
\usepackage{acro}
\usepackage{graphicx}
\usepackage{textcomp}
\usepackage{xcolor}
\usepackage{float}

\input{acronyms}
\def\BibTeX{{\rm B\kern-.05em{\sc i\kern-.025em b}\kern-.08em
    T\kern-.1667em\lower.7ex\hbox{E}\kern-.125emX}}

\begin{document}

\title{Log-Mu Fading Process: Second-Order Statistics for Diversity-Combining Techniques \\

\thanks{This work was partially supported by the São Paulo Research Foundation, FAPESP, (2024/20097-3).}
}

\author{\IEEEauthorblockN{Godfred Kumi Tenkorang }
\IEEEauthorblockA{Department of Communications \\ Faculty of Electrical and Computer Engineering \\
State University of Campinas\\
Campinas, Brazil \\
g218490@dac.unicamp.br}
\and
\IEEEauthorblockN{Michel Daoud Yacoub}
\IEEEauthorblockA{Department of Communications \\ Faculty of Electrical and Computer Engineering \\ 
State University of Campinas\\
Campinas, Brazil \\
mdyacoub@unicamp.br}

}

\maketitle

\begin{abstract}

This paper derives second-order statistics for diversity-combining techniques over Log-mu fading channels. Closed-form expressions for the \ac{lcr} and \ac{afd} are derived for \ac{psc}, while exact multidimensional integral expressions are obtained for \ac{egc} and \ac{mrc}. The analysis considers $M$ unbalanced, independent, and non-identically distributed (i.n.i.d.) Log-mu fading channels. Monte Carlo simulations are conducted to validate the theoretical results, demonstrating excellent agreement and confirming the accuracy of the proposed expressions. 

\end{abstract}


\section{Introduction}
\IEEEPARstart{I}n wireless communication systems, the performance of received signals is significantly impacted by fading, which arises primarily from two key phenomena, namely shadowing and multipath propagation. Shadowing, often referred to as long-term fading, is caused by obstacles blocking the line-of-sight (LoS) path, leading to slow variations in signal strength over large distances. In contrast, multipath propagation, also known as short-term fading, arises from the diffraction, reflection, and scattering of transmitted signals, causing rapid fluctuations in signal amplitude and phase. A wide range of statistical models have been proposed and extensively studied in the literature to characterize these fading effects.

Among these, $\alpha$-$\mu$ arose to explicitly incorporate the concept of the non-linear effects of the propagation medium, rendering it more versatile to characterize short-term fading. Due to its non-linear structure and mathematical tractability, the authors in \cite{Anjosα-µforComposite} investigated its suitability for composite fading scenarios, which, implicitly, is a non-linear phenomenon. Their study compared the statistical fitting accuracy of the $\alpha$–$\mu$ model with well-established composite fading models, Nakagami–Lognormal, Generalized-$K$, and Fisher–Snedecor, based on field measurements at 1.8 GHz. Their results demonstrated that the $\alpha$–$\mu$ model either outperformed these benchmarks or achieved comparable performance. Motivated by these findings and the growing need for models that capture non-linear propagation effects with simpler mathematical formulations, the authors in \cite{tenkorang2025logmu} recently introduced the Log–$\mu$ fading model, which is revisited in section \ref{sec2} of this paper.

As is widely known, the same phenomenon that provokes fading, i.e., multipath, can also be used to counteract its deleterious effects. Hence, diversity combining is a crucial technique in wireless communication systems to improve performance. Evaluating the performance of these techniques relies on key first- and second-order statistics, for instance, outage probability, \ac{lcr}, \ac{afd}, the latter two providing insight into the temporal dynamics of fading processes. These metrics have been extensively studied for classical fading models \cite{Zeng2024,Stefanovic2022,SaucoGallardo2017, DaCosta2007}, however, their characterization under the recently proposed Log–$\mu$ fading channel has not yet been addressed.

In this paper, we derive the expressions of the \ac{lcr} and \ac{afd} for the three conventional diversity-combining schemes, namely \ac{psc}, \ac{egc}, and \ac{mrc} techniques, in the presence of $M$ unbalanced, independent, and non-identically distributed (i.n.i.d.) Log–$\mu$ fading channels. For the \ac{psc} scheme, closed-form expressions are obtained, whereas the \ac{egc} and \ac{mrc} statistics are expressed in the form of multidimensional integrals. The analytical derivations are validated through Monte Carlo simulations, confirming the accuracy of the proposed results.

\section{The Log-$\mu$ Fading Model Revisited}
\label{sec2}
In this section, we revisit the Log-$\mu$ fading model, originally introduced in \cite{tenkorang2025logmu}. This model considers that the received signal at the receptor arises from the superposition of $\mu$ clusters of multipath components propagating through a non-homogeneous environment. The envelope of the resulting signal is described as a non-linear transformation of the modulus of the sum of these multipath waves, where the non-linearity of the propagation medium is captured by a logarithmic function. Accordingly, the normalized envelope, $P_i = R_i /\hat{r}_i$, at the $i$-th branch, $i = 1, 2, ...,M$, can be expressed as

\begin{equation}
g(P_i) =  \log\left( \left( P_i + (1 - s_i)^{1/\alpha_i}\right)^{\alpha_i} + s_i \right) = \sum_{l=1}^{\mu_i} ({X}^2_{il} + {Y}^2_{il}),
 \label{eq1}
\end{equation}
where $\alpha_i > 0$ and $s_i \in \mathbb{R}$ are arbitrary parameters, $\hat{r}_i$ is a scale parameter, and ${X}_{il}$ and ${Y}_{il}$ are mutually independent Gaussian processes with zero mean and equal variance. The \ac{pdf} of the normalized envelope, $f_{P_i}(\rho_i)$, is given as \cite{tenkorang2025logmu} 

\begin{equation}\begin{aligned}
 f_{P_i}(\rho_i) = \frac{ \alpha_i {({\rho_i + (1-s_i)^{1/ \alpha_i}}) ^ {\alpha_i -1}}\mu_i^{\mu_i} }{\Gamma(\mu_i) (({\rho_i + (1-s_i)^{1/ \alpha_i}})^{\alpha_i} +s_i)^{\mu_i +1}} \\ \times \log(({\rho_i + (1-s_i)^{1/ \alpha_i}})^{\alpha_i} + s_i)^{\mu_i-1}, 
 \label{eq2}
 \end{aligned}
\end{equation}
where $\Gamma(u) = \int_{0}^{\infty} {x^{u -1} exp(-x) dx} $ is the Gamma function. The parameter $\alpha_i$ is unrestricted for $s_i \leq 1$; however, for $s_i > 1$, $\alpha_i$ must be restricted to odd integers to ensure real-valued expressions. 

The corresponding \ac{cdf} of the normalized envelope, $F_{P_i}(\rho_i)$ is obtained in a closed-form as 

\begin{equation}
 F_{P_i}(\rho_i) = 1-\frac{\Gamma\left( \mu_i, \mu_i \log\left( \left( {\rho_i}+ (1 - s_i)^{1/\alpha_i}\right)^{\alpha_i} + s_i \right) \right) }{ \Gamma(\mu_i) },
 \label{eq3}
\end{equation}
where $\Gamma(u,v) = \int_{v}^{\infty} x^{u-1}\exp(-x)dx$ is the incomplete Gamma function.

A distinctive feature of the Log-$\mu$ fading model is its ability to exhibit bimodal characteristics for certain parameter configurations, particularly when $\alpha > 1$ and $s > 1$.

\section{Level Crossing Rate  and Average Fade Duration}

The \ac{lcr} is defined as the average number of times a fading signal crosses a given signal level $\rho$ within a certain period of time. The \ac{lcr} in the positive direction can be obtained by 

\begin{equation}
N_{P} (\rho) = \int_0^{\infty} \dot{\rho} f_{P,\dot{P}}(\rho,\dot{\rho}) d\dot{\rho}, 
\label{eq4}
\end{equation}
where $\dot{P}$ is the time derivative of $P$, and $f_{P,\dot{P}}(\cdot,\cdot)$ is the joint \ac{pdf} of $P$ and $\dot{P}$. 

The \ac{afd} represents the average time the signal remains below the threshold $\rho$, and it can be expressed as
\begin{equation}
T_{P}(\rho) = \frac{F_{P}(\rho)}{N_P(\rho)},
\label{eq5}
\end{equation}
where $F_{P}(\cdot)$ is the \ac{cdf}.

\subsection{No Diversity ($M = 1$)}

These second-order statistics can be derived using any suitable relation between the envelope of the Log-$\mu$ model and the envelope of another well-characterized fading model. From \eqref{eq1}, it can be observed that the non-linear transformation applied to the received signal relates the Log-$\mu$ envelope to the Nakagami-$m$ model through the expression

\begin{equation}
g(P_i) = {P}_{i,\,N}^{2},
\label{eq6}
\end{equation}
where ${P_i}$ and ${P}_{i,\,N}$ represent the normalized envelopes of the Log-$\mu$ and Nakagami-$m$ fading models, respectively.

By differentiating both sides of \eqref{eq6} with respect to time, we obtain

\begin{equation}\begin{aligned}
\dot{{P_i}} = \frac{2 \left(g({P_i})\right)^{0.5} \dot{{P}}_{i,\,N} }{g'({P_i})}, \label{eq7}
\end{aligned}\end{equation}
where $g'({P_i}) = dg({P_i})/d{P_i}$. As presented in \cite{yacoub1999higher}, the time derivative $\dot{{P}}_{i,\,N}$ of the Nakagami-$m$ envelope follows a zero-mean Gaussian distribution with variance

\begin{equation}
{\sigma}_{\dot{P}_i,\,N}^2  =  \left(\frac{\omega}{2}\right)^2 \left(\frac{1}{\mu_i}\right), 
 \label{eq8}
\end{equation}
where $\omega$ is the maximum Doppler shift in radians per second. Applying the standard procedure for transformation of random variables to \eqref{eq7}, $f_{\dot{P}_i|P_i}(\cdot|\cdot)$ is also found to be Gaussian with zero mean and variance

\begin{equation}\begin{aligned}
{\sigma}_{\dot{P}_i}^2 = \frac{g({P_i}) \omega^2}{\left(g'({P_i})\right)^2 \mu_i }. 
 \label{eq9}
\end{aligned}\end{equation}

The joint \ac{pdf} of the envelope and its time derivative is then given as $f_{\dot{P_i},P_i}(\dot{\rho}_i,\rho_i) = f_{\dot{P}_i|P_i}(\dot{\rho}_i|\rho_i) f_{P_i}(\rho_i)$ and by substituting this into \eqref{eq4}, the \ac{lcr} is obtained in closed-form as

\begin{equation}\begin{aligned}
N_{P_i}(\rho) = \frac{{\sigma}_{\dot{P}_i} f_{P_i}({\rho})}{\sqrt{2\pi}}.
 \label{eq10}
 \end{aligned}\end{equation}

More explicitly, the \ac{lcr} for the Log-$\mu$ fading envelope is expressed as

\begin{equation}\begin{aligned}
N_{P_i}(\rho)  = \frac{\omega \mu_i^{\mu_i -0.5} \left(\log\left( \left( {\rho} + (1 - s_i)^{1/\alpha_i}\right)^{\alpha_i} + s_i \right)\right)^{\mu_i - 0.5}}{\sqrt{2 \pi} {\Gamma(\mu_i) \left( \left( {\rho} + (1 - s_i)^{1/\alpha_i}\right)^{\alpha_i} + s_i \right)^{\mu_i}}}.
\label{eq11}
\end{aligned}\end{equation}

The \ac{afd} can be derived in a closed-form by substituting \eqref{eq3} and \eqref{eq11} into \eqref{eq5}, yielding

\begin{equation}\begin{aligned}
T_{P_i}(\rho)  = {\sqrt{2 \pi} { \left( \left( {\rho} + (1 - s_i)^{1/\alpha_i}\right)^{\alpha_i} + s_i \right)^{\mu_i}}} \\ & \hspace{-18em} \times \frac{\left(\Gamma(\mu_i)- \Gamma\left( \mu_i, \mu_i \log\left( \left( {\rho} + (1 - s_i)^{1/\alpha_i}\right)^{\alpha_i} + s_i \right) \right)\right)}{{\omega \mu_i^{\mu_i -0.5} \left(\log\left( \left( {\rho} + (1 - s_i)^{1/\alpha_i}\right)^{\alpha_i} + s_i \right)\right)^{\mu_i - 0.5}}}.
\label{eq12}
\end{aligned}\end{equation}

\subsection{Pure Selection Combining}

In \ac{psc}, the receiver continuously monitors all branches and selects the one with the highest instantaneous signal envelope. Consequently, the output envelope $P$ at the combiner is defined as

\begin{equation}
 P = \max\limits_{\substack{i = 1, \dots, M}} \{P_{i}\}.
 \label{eq13}
\end{equation}

In \cite{YangLin2007}, a general formulation for the \ac{lcr} of \ac{psc} system over $M$ independent fading channels was derived. Based on this formulation, the \ac{lcr} for \ac{psc} over Log-$\mu$ fading channels is obtained as 

\begin{equation}
N_P(\rho) = \sum_{i=1}^{M} N_{P_i}(\rho) \prod_{\substack{j = 1 \\ j \neq i}}^{M} {F_{P_j}(\rho)}, 
 \label{eq14}
\end{equation}
where $F_{P_j}$ and $N_{P_i}$ are given by \eqref{eq3} and \eqref{eq11}, respectively.

A general formulation for the \ac{afd} of \ac{psc} system over $M$ independent fading channels has also been derived in \cite{DaCosta2007}. It is expressed as 

\begin{equation}
T^{-1}_{P} (\rho) = \sum_{i=1}^{M}  {T^{-1}_{P_i}(\rho)}, 
 \label{eq15}
\end{equation}
where ${T_{P_i}}$ is given in our case by \eqref{eq12}.

\subsection{Equal Gain Combining}

In \ac{egc}, the signals received with envelopes $\rho_i$ are co-phased and coherently added. Taking into account the resultant noise power at the combiner output, the overall output envelope $P$ and its time derivative $\dot{P}$ are given as

\begin{equation}
P = \frac{1}{\sqrt{M}}\sum_{i=1}^{M} P_i
\quad \text{and} \quad
\dot{P} = \frac{1}{\sqrt{M}}\sum_{i=1}^{M} \dot{P_i} .
\label{eq16}
\end{equation}

Following a similar procedure to that in \cite{Brennan1957}, where the distribution of the sum of $M$ independent Rayleigh signals was derived, we obtain the distribution of the Log-$\mu$ signals over the $M$-dimensional volume bounded by the hyperplane $\sqrt{M} \rho = \sum_{i=1}^{M} \rho_i$ as 

\begin{equation}
\begin{aligned}
F_{P}(\rho) = \int_0^{\sqrt{M}\rho} \int_0^{\sqrt{M}\rho - \rho_M} \dots \int_0^{\sqrt{M}\rho - \sum_{i=3}^{M} {\rho_i}} \int_0^{\sqrt{M}\rho - \sum_{i=2}^{M} {\rho_i}} \\ \times \prod_{i=1}^{M} f_{P_i}(\rho_i) d\rho_1 d\rho_2 ...d\rho_{M-1} d\rho_{M},
\label{eq17}
\end{aligned}
\end{equation}
where $f_{P_i}(\rho_i)$ is given by \eqref{eq2}. Differentiating \eqref{eq17} with respect to $\rho$ yields the envelope \ac{pdf} $f_{P}(\rho)$, and using Bayes' rule, the joint distribution $f_{P,\dot{P}}(\rho, \dot{\rho})$ is found as \eqref{eq18}.
The joint density function $f( \rho_1, \rho_2, \dots, \rho_M, \dot{\rho})$ is expressed as

\begin{figure*}[!t]
\begin{equation}
\begin{aligned}
f_{P,\dot{P}}(\rho,\dot{\rho}) = \sqrt{M} \int_0^{\sqrt{M}\rho} \int_0^{\sqrt{M}\rho - \rho_M} \dots \int_0^{\sqrt{M}\rho - \sum_{i=3}^{M} \rho_i} \\ & 
\hspace{-2.5em} \times f_{P_1,P_2,\cdots,P_M,\dot{P}}\left( \left(\sqrt{M}\rho - \sum_{i=2}^{M} \rho_i\right), \rho_2, \dots, \rho_M, \dot{\rho} \right) \, d\rho_2 \dots d\rho_{M-1} d\rho_M.
\end{aligned}
\label{eq18}
\end{equation}
 \rule{\linewidth}{0.4pt}
\end{figure*}


\begin{equation}
\begin{aligned}
f_{P_1, \dots, P_M, \dot{P}} \left( \rho_1, \dots, \rho_M, \dot{\rho} \right) = f_{\dot{P} | P_1, \dots, P_M} \left( \dot{\rho} | \rho_1, \dots, \rho_M \right) \\ \times f_{P_1, \dots, P_M} \left( \rho_1, \dots, \rho_M \right)
\end{aligned}
\label{eq19}
\end{equation} 
where $f\left(\dot{\rho}|\rho_1, \dots, \rho_M \right)$ is a zero-mean Gaussian distribution with variance ${\sigma}_{\dot{P}}^2 = \sum_{i=1}^{M} {{\sigma}_{\dot{P}_i}^2}/M$, and $f_{P_1, \dots, P_M} \left( \rho_1, \dots, \rho_M \right) =  \prod_{i=1}^{M} f_{P_i}(\rho_i)$. By substituting \eqref{eq19} into \eqref{eq18} and subsequently incorporating the result into \eqref{eq4}, we obtain the expression for the \ac{lcr} of the \ac{egc} system, as presented in \eqref{eq20}. The corresponding \ac{afd} is then obtained directly with \eqref{eq20}, \eqref{eq17}, and \eqref{eq4}.


\begin{figure*}[!t]
\begin{equation}
\begin{aligned}
N_{P}(\rho) = \frac{\omega}{\sqrt{2\pi}} \int_0^{\sqrt{M}\rho} \int_0^{\sqrt{M}\rho - \rho_M} \dots \int_0^{\sqrt{M}\rho - \sum_{i=3}^{M} \rho_i} \sqrt{\frac{g\left(\sqrt{M} \rho -\sum_{i=2}^{M} \rho_i \right)}{\left(g'\left(\sqrt{M} \rho -\sum_{i=2}^{M} \rho_i \right)\right)^2 \mu_1} + \sum_{i=2}^{M}\frac{g(\rho_i)}{(g'(\rho_i))^2 \mu_i}}  \\ \times f_{P_1}\left(\sqrt{M}\rho - \sum_{i=2}^{M} \rho_i\right) \prod_{i=2}^{M} f_{P_i}(\rho_i) d\rho_2 \dots d\rho_{M-1} d\rho_M.
\end{aligned}
\label{eq20}
\end{equation}
 \rule{\linewidth}{0.4pt}
\end{figure*}

\subsection{Maximal Ratio Combining}
In \ac{mrc}, the received signals are co-phased, each signal is amplified appropriately to achieve optimal combining. The resulting signals are then summed, such that the combiner output envelope $P$ and its time derivative $\dot{P}$ are given by

\begin{equation}
P = {\sqrt{\sum_{i=1}^{M} P_i^2}}
\quad \text{and} \quad
\dot{P} = \sum_{i=1}^{M} \frac{P_i}{P}\dot{P_i} .
\label{eq21}
\end{equation}

The \ac{mrc} analysis follows the same general procedure as that of \ac{egc}. However, in the case of \ac{mrc}, the distribution is derived over a $M$-dimensional volume bounded by the hyperplane $\rho^2 = \sum_{i=1}^{M} \rho_i^2$. The resulting variance of $\dot{P}$ can be expressed as ${\sigma}_{\dot{P}}^2 = \sum_{i=1}^{M} {{P_i ^2}{\sigma}_{\dot{P}_i}^2}/P^2$. The \ac{cdf} is expressed as 

\begin{equation}
\begin{aligned}
F_{P}(\rho) = \int_0^{\rho} \int_0^{\sqrt{\rho^2 - \rho_M^2}} \dots \int_0^{\sqrt{\rho^2 - \sum_{i=3}^{M} {\rho_i^2}}} \int_0^{\sqrt{\rho^2 - \sum_{i=2}^{M} {\rho_i^2}}} \\ \times \prod_{i=1}^{M} f_{P_i}(\rho_i) d\rho_1 d\rho_2 ...d\rho_{M-1} d\rho_{M}.
\label{eq22}
\end{aligned}
\end{equation}

The joint \ac{pdf} $f_{P,\dot{P}}(\cdot,\cdot)$ and the \ac{lcr} $N_{P}(\cdot)$ are expressed by \eqref{eq23} and \eqref{eq24}, respectively. The corresponding \ac{afd} is then obtained directly with \eqref{eq24}, \eqref{eq22}, and \eqref{eq4}.

\begin{figure*}[!t]
\begin{equation}
\begin{aligned}
f_{P,\dot{P}}(\rho,\dot{\rho}) = \int_0^{\rho} \int_0^{\sqrt{\rho^2 - \rho_M^2}} \dots \int_0^{\sqrt{\rho^2 - \sum_{i=3}^{M} {\rho_i^2}}} \frac{\rho}{\sqrt{\rho^2-\sum_{i=2}^{M} \rho_i^2}} \\ & 
\hspace{-5.5em} \times f_{P_1,P_2,\cdots,P_M,\dot{P}}\left( \left({\sqrt{\rho^2-\sum_{i=2}^{M} \rho_i^2}}\right), \rho_2, \dots, \rho_M, \dot{\rho} \right) \, d\rho_2 \dots d\rho_{M-1} d\rho_M.
\end{aligned}
\label{eq23}
\end{equation}
 \rule{\linewidth}{0.4pt}
\end{figure*}

\begin{figure*}[!t]
\begin{equation}
\begin{aligned}
N_{P}(\rho) = \frac{\omega}{\sqrt{2\pi}} \int_0^{\rho} \int_0^{\sqrt{\rho^2 - \rho_M^2}} \dots \int_0^{\sqrt{\rho^2 - \sum_{i=3}^{M} {\rho_i^2}}} \frac{1}{\sqrt{\rho^2-\sum_{i=2}^{M} \rho_i^2}} \sqrt{\frac{g\left({\sqrt{\rho^2-\sum_{i=2}^{M} \rho_i^2}}\right) \left({{\rho^2-\sum_{i=2}^{M} \rho_i^2}}\right)}{\left(g'\left({\sqrt{\rho^2-\sum_{i=2}^{M} \rho_i^2}}\right)\right)^2 \mu_1} + \sum_{i=2}^{M}\frac{g(\rho_i) \rho_i^2}{(g'(\rho_i))^2 \mu_i}}  \\ \times f_{P_1}\left({\sqrt{\rho^2-\sum_{i=2}^{M} \rho_i^2}}\right) \prod_{i=2}^{M} f_{P_i}(\rho_i) d\rho_2 \dots d\rho_{M-1} d\rho_M.
\end{aligned}
\label{eq24}
\end{equation}
 \rule{\linewidth}{0.4pt}
\end{figure*}

\section{Results}

This section presents numerical results to illustrate the derived expressions and validate their accuracy through Monte Carlo simulations. All diversity branches are assumed to be balanced, independent, and identically distributed (i.i.d.), and curves corresponding to single-branch transmission (no diversity) are omitted for clarity.  

\begin{figure}[hbt]
    \centering
    \includegraphics[width=0.8\linewidth]{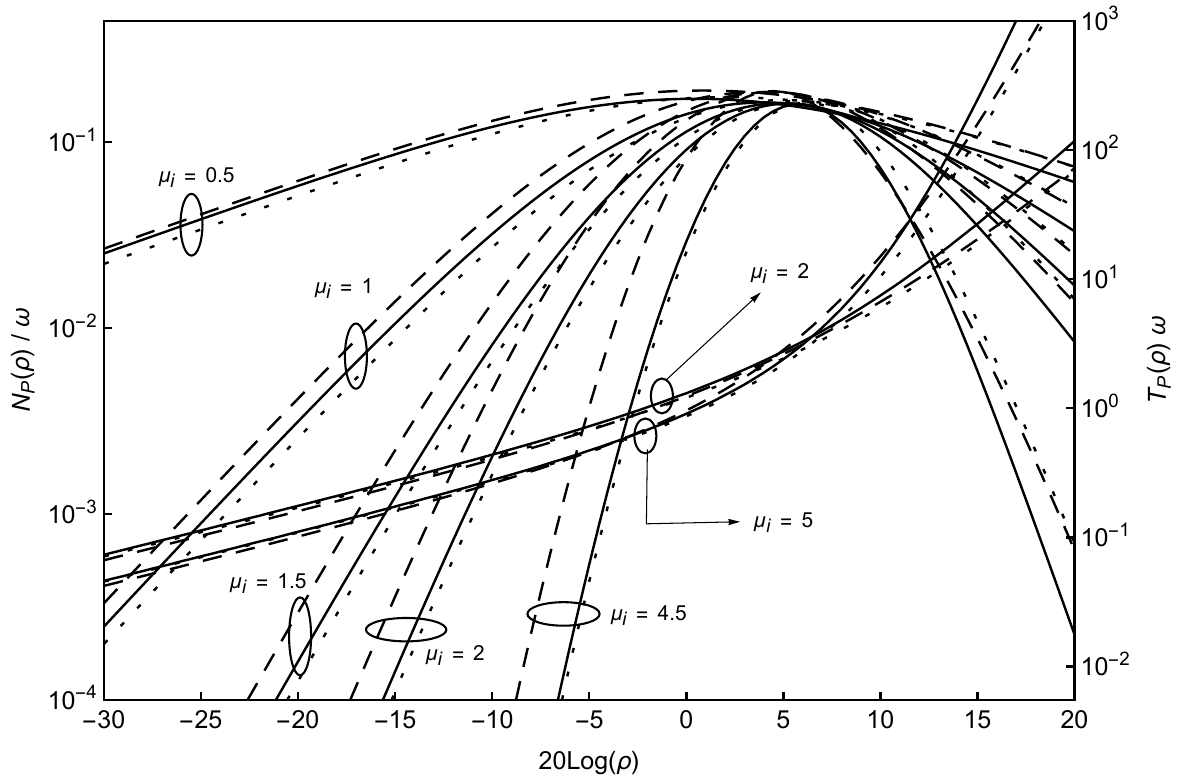}
    \caption{\ac{lcr} and \ac{afd} of \ac{psc}, \ac{egc}, and \ac{mrc} techniques over Log-$\mu$ fading channels (dashed lines $\rightarrow$ \ac{psc}, solid lines $\rightarrow$ \ac{egc}, dotted lines $\rightarrow$ \ac{mrc}, $M = 2$, $s_i = 1$, $\alpha_i = 1.5$, and varying $\mu_i $). \label{M2_lcr_afd_varyingmu} }
\end{figure}

\begin{figure}[hbt]
    \centering
    \includegraphics[width=0.8\linewidth]{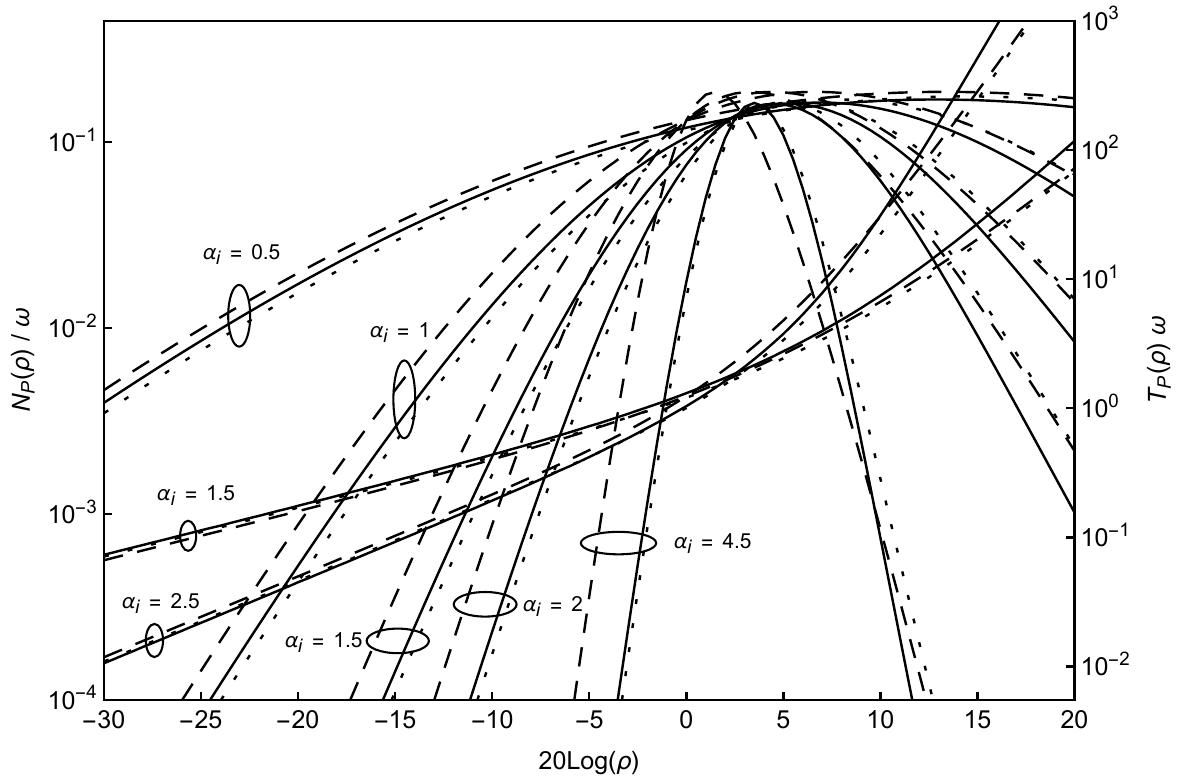}
    \caption{\ac{lcr} and \ac{afd} of \ac{psc}, \ac{egc}, and \ac{mrc} techniques over Log-$\mu$ fading channels (dashed lines $\rightarrow$ \ac{psc}, solid lines $\rightarrow$ \ac{egc}, dotted lines $\rightarrow$ \ac{mrc}, $M = 2$, $s_i = 1$, $\mu_i = 2$, and varying $\alpha_i $).\label{M2_lcr_afd_varyingalpha} }
\end{figure}

\begin{figure}[hbt]
    \centering
    \includegraphics[width=0.8\linewidth]{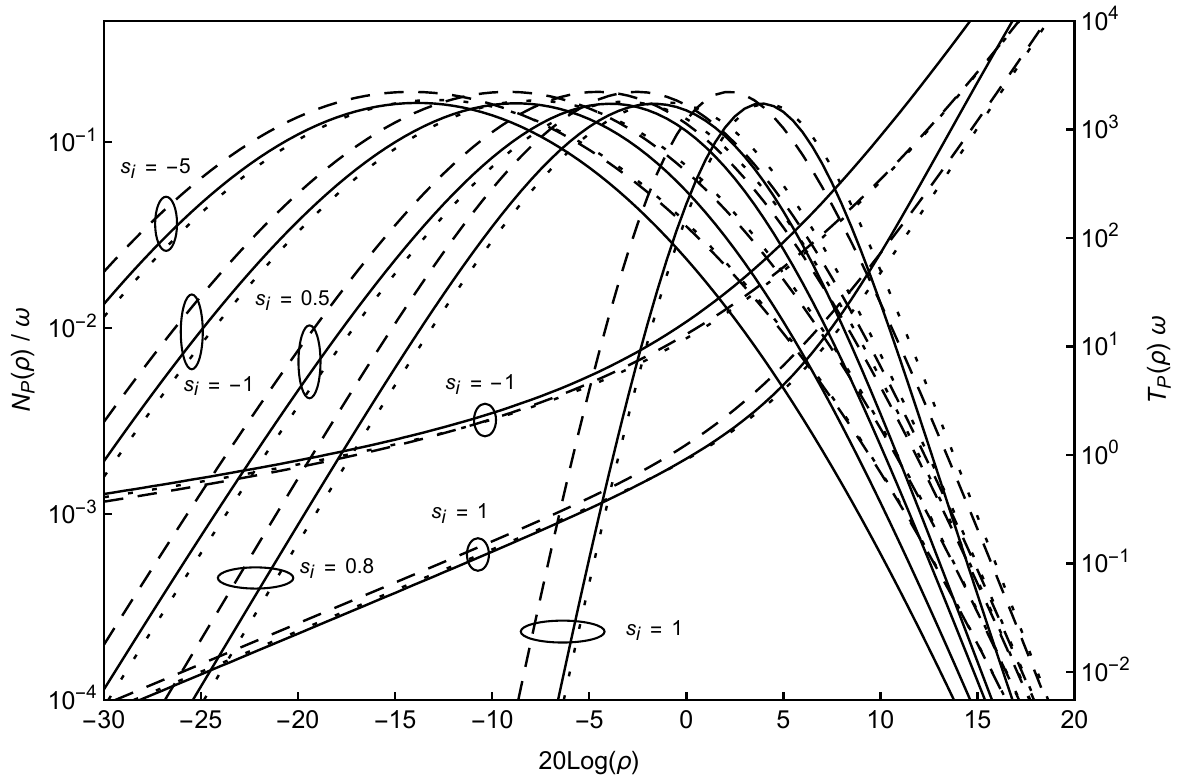}
    \caption{\ac{lcr} and \ac{afd} of \ac{psc}, \ac{egc}, and \ac{mrc} techniques over Log-$\mu$ fading channels (dashed lines $\rightarrow$ \ac{psc}, solid lines $\rightarrow$ \ac{egc}, dotted lines $\rightarrow$ \ac{mrc}, $M = 2$, $\alpha_i = 3$, $\mu_i = 2$, and varying $s_i$).\label{M2_lcr_afd_varyings} }
\end{figure}

\begin{figure}[hbt]
    \centering
    \includegraphics[width=0.8\linewidth]{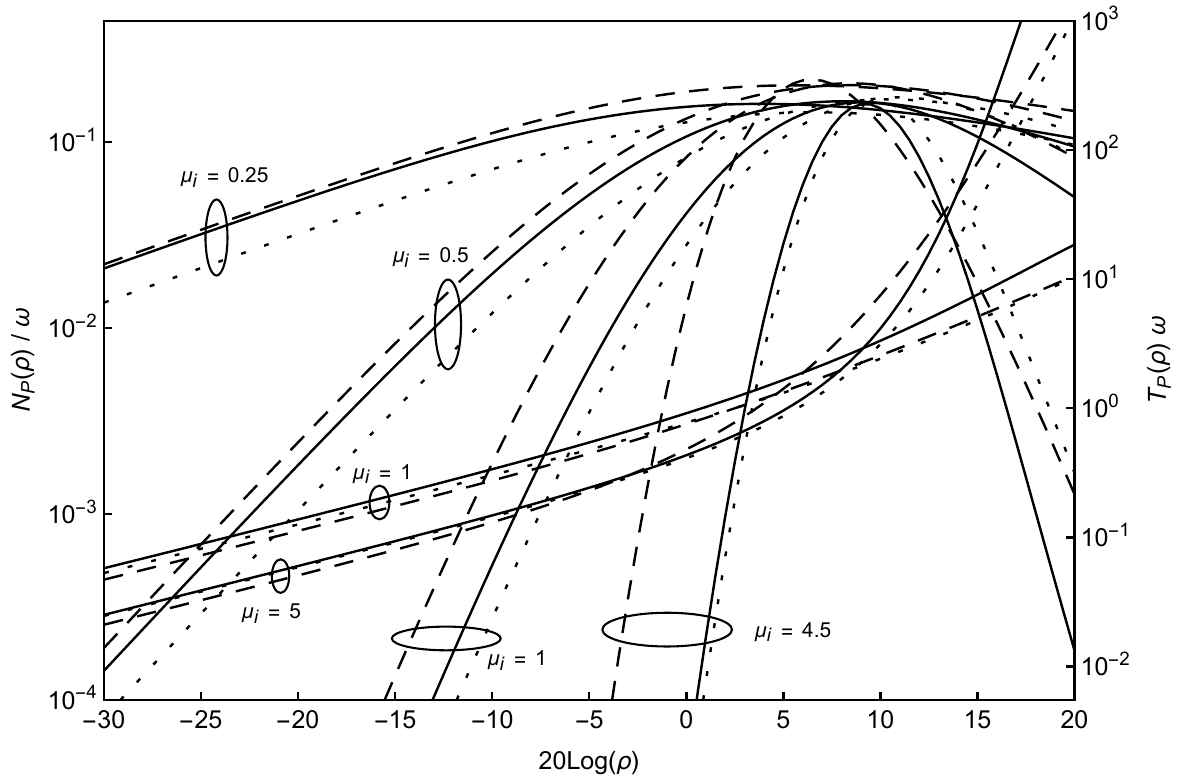}
    \caption{\ac{lcr} and \ac{afd} of \ac{psc}, \ac{egc}, and \ac{mrc} techniques over Log-$\mu$ fading channels (dashed lines $\rightarrow$ \ac{psc}, solid lines $\rightarrow$ \ac{egc}, dotted lines $\rightarrow$ \ac{mrc}, $M = 4$, $s_i = 1$, $\alpha_i = 1.5$, and varying $\mu_i $).\label{M4_lcr_afd_varyingmu} }
\end{figure}

Fig. \ref{M2_lcr_afd_varyingmu} illustrates the behavior of the \ac{lcr} and \ac{afd} for \ac{psc}, \ac{egc}, and \ac{mrc} with $M = 2$ diversity branches, considering fixed $s_i$ and $\alpha_i$ while varying $\mu_i$.  
Fig. \ref{M2_lcr_afd_varyingalpha} presents results for the same number of branches with fixed $s_i$ and $\mu_i$, varying $\alpha_i$.  
To further analyze the effect of the arbitrary parameters, Fig. \ref{M2_lcr_afd_varyings} shows plots for $M = 2$ diversity branches with fixed $\alpha_i$ and $\mu_i$ while varying $s_i$.  
The analysis is then extended in Fig. \ref{M4_lcr_afd_varyingmu} to a four-branch diversity system ($M = 4$) with fixed $s_i$ and $\alpha_i$ and varying $\mu_i$.  

A comparison of Fig. \ref{M2_lcr_afd_varyingmu} and Fig. \ref{M4_lcr_afd_varyingmu} shows that, under identical fading conditions, increasing the number of diversity branches enhances the output signal, leading to a notable shift in crossing behavior; lower thresholds are crossed less frequently, while higher thresholds are crossed more frequently, reflecting improved link reliability and reduced occurrence of deep fading.  

\begin{figure}[hbt]
    \centering
    \includegraphics[width=0.8\linewidth]{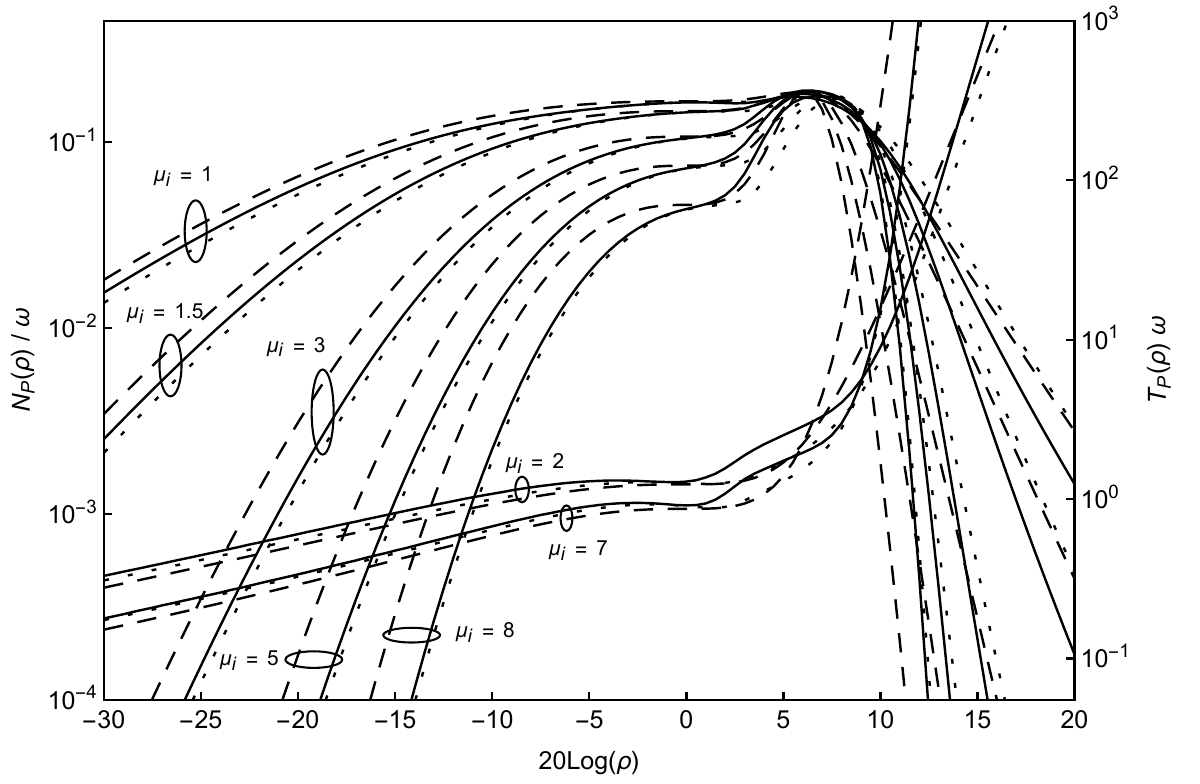}
    \caption{\ac{lcr} and \ac{afd} of \ac{psc}, \ac{egc}, and \ac{mrc} techniques over Log-$\mu$ fading channels (dashed lines $\rightarrow$ \ac{psc}, solid lines $\rightarrow$ \ac{egc}, dotted lines $\rightarrow$ \ac{mrc}, $M = 2$, $s_i = 2$, $\alpha_i = 3$, and varying $\mu_i $ (bimodal scenario)).\label{M2bimodal_lcr_afd_varyingmu} }
\end{figure}

Knowing that the Log-$\mu$ \ac{pdf} can exhibit bimodal behavior for certain parameter configurations, Fig. \ref{M2bimodal_lcr_afd_varyingmu} provides additional insight into this scenario by plotting the \ac{lcr} and \ac{afd} for such parameter sets. This highlights the flexibility of the Log-$\mu$ model in capturing non-standard fading characteristics.  

\begin{figure}[hbt]
    \centering
    \includegraphics[width=0.8\linewidth]{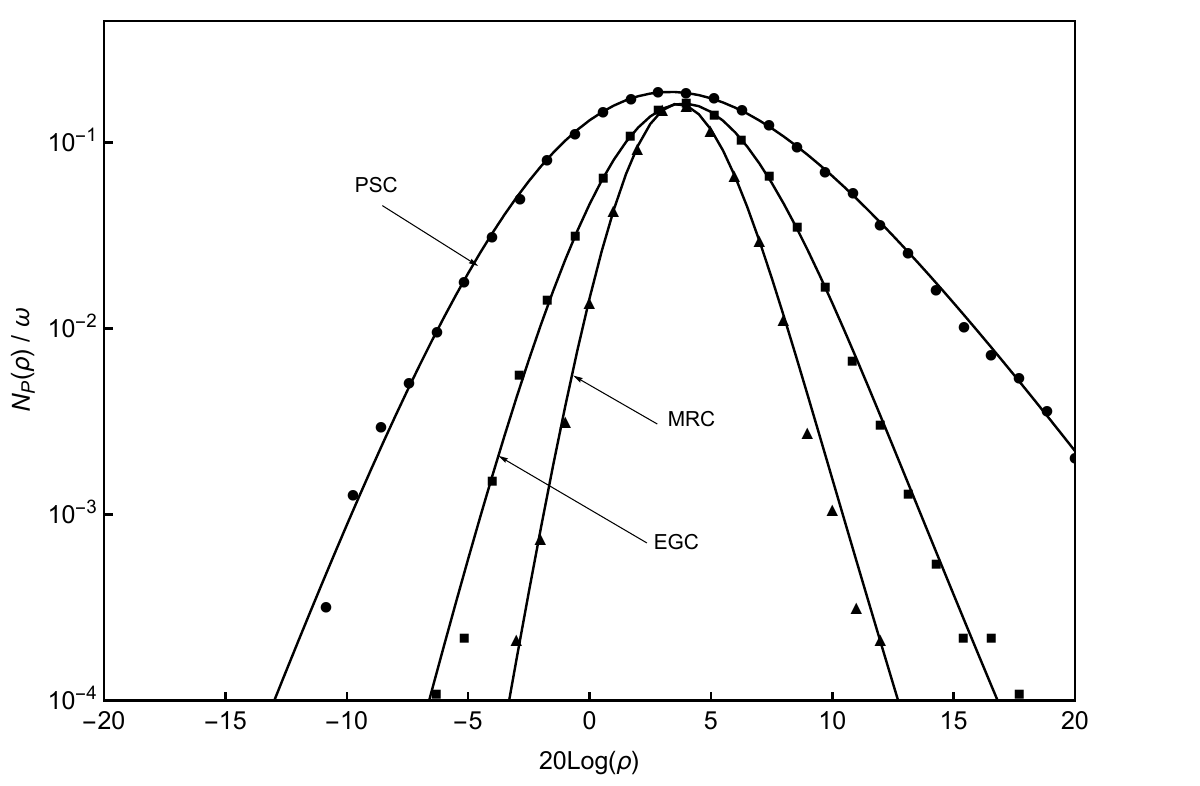}
    \caption{Comparison between simulated and theoretical curves (\ac{psc}: $M = 2$, $s_i = 1$, $\alpha_i = 2$, $\mu_i = 2$ , \ac{egc}: $M = 2$, $s_i = 1$, $\alpha_i = 3$, $\mu_i = 2$, \ac{mrc}: $M = 2$, $s_i = 1$, $\alpha_i = 4.5$, $\mu_i = 2$). \label{montecarlo} }
\end{figure}

Across all scenarios and parameter configurations, \ac{psc} consistently demonstrates the lowest performance, whereas \ac{egc} and \ac{mrc} exhibit comparable and superior performance levels, with \ac{mrc} generally achieving the best overall performance. Finally, Fig. \ref{montecarlo} compares the analytical results with Monte Carlo simulations for $M = 2$ branches, showing excellent agreement and validating the accuracy of the derived expressions.

\section{Conclusion}

In this paper, we derived expressions for the \ac{lcr} and \ac{afd} of \ac{psc}, \ac{egc}, and \ac{mrc} techniques over $M$ unbalanced, independent, and non-identically distributed (i.n.i.d.) Log-$\mu$ fading channels. For the \ac{psc} scheme, exact closed-form expressions for both metrics were obtained, while in the \ac{egc} and \ac{mrc} cases, the solutions were expressed as multidimensional integrals. The analytical results were validated through Monte Carlo simulations, demonstrating excellent agreement and confirming the accuracy of the proposed expressions.


\bibliographystyle{IEEEtran}
\bibliography{sources.bib}


\vspace{12pt}

\end{document}

%% file: acronyms.tex
\DeclareAcronym{pdf}{
	short = PDF,
	long  = probability density function,
}
\DeclareAcronym{los}{
	short = LoS,
	long  = line-of-sight,
}
\DeclareAcronym{nlos}{
	short = nLoS,
	long  = non-line-of-sight,
}
\DeclareAcronym{afd}{
	short =$\text{AFD}$,
	long  =average fading duration,
}

\DeclareAcronym{rv}{
	short = RV,
	long  = random variable,
}
\DeclareAcronym{lcr}{
	short =$\text{LCR}$,
	long  =level crossing rate,
}

\DeclareAcronym{tx}{
	short = Tx,
	long  = transmitter,
}
\DeclareAcronym{rx}{
	short = Rx,
	long  = receiver,
}

\DeclareAcronym{cdf}{
	short = CDF,
	long  = cumulative distribution function,
}
\DeclareAcronym{snr}{
	short = SNR,
	long  = signal-to-noise ratio,
}
\DeclareAcronym{}{
	short =$\text{LCR}_\text{t}$,
	long  = level crossing rate,
}

\DeclareAcronym{nmse}{
	short = NMSE,
	long  = normalized mean square error,
}

\DeclareAcronym{psc}{
	short = PSC,
	long  = pure selection combining,
}

\DeclareAcronym{egc}{
	short = EGC,
	long  = equal gain combining,
}

\DeclareAcronym{mrc}{
	short = MRC,
	long  = maximal ratio combining,
}
\DeclareAcronym{mod-ks}{
	short = modified K-S, 
	long = modified Kolmogorov-Smirnov test,
}

